\documentclass[sigconf]{acmart}
\copyrightyear{2026}
\acmYear{2026}
\setcopyright{cc}

\acmConference[CIKM '26]
{Proceedings of the 35th ACM International Conference on Information and Knowledge Management}
{November 7--11, 2026}
{Rome, Italy.}

\acmBooktitle{Proceedings of the 35th ACM International Conference on Information and Knowledge Management (CIKM '26), November 7--11, 2026, Rome, Italy}

\setcctype{by}
\acmISBN{979-8-4007-2539-5/2026/11}
\acmDOI{10.1145/3799682.3839965}
\newcommand{\best}[1]{\textbf{#1}}
\newcommand{\ndcg}{nDCG@10}

\usepackage{algorithm}
\usepackage{algpseudocode}
\usepackage{amsmath}
\usepackage{amsfonts}
\usepackage{booktabs}
\usepackage{balance}
\usepackage{microtype}
\usepackage{amsmath}
\usepackage{graphicx}
\usepackage{pgfplots}
\usepackage[utf8]{inputenc}  % Most likely already there
\usepackage[T1]{fontenc}
\usepackage{textcomp}
\usepackage{enumitem}
\usepackage{multirow}
\usepackage{makecell}
\usepackage[most]{tcolorbox}
\usepackage{xcolor}

\usepackage{algorithm}
\usepackage{algpseudocode}
\usepackage{amsmath}      % for \begin{cases}, \dfrac, \mathbb

\usepackage{amssymb}      % for \mathbb{I} (indicator)

\author{Hai Son Le}
\orcid{0009-0003-2240-0451}
\affiliation{%
  \institution{Toronto Metropolitan University}
  \city{Toronto}
  \country{Canada}
  }

\author{Amin Bigdeli}
\orcid{0009-0003-8977-9312}
\affiliation{%
  \institution{University of Waterloo}
  \city{Toronto}
  \country{Canada}
  }

\author{Shirin Seyedsalehi}
\orcid{0009-0006-1467-1659}

\affiliation{%
  \institution{Toronto Metropolitan University}
  \city{Toronto}
  \country{Canada}
}

\author{Morteza Zihayat}
 \orcid{0000-0002-1144-7364}
\affiliation{%
  \institution{Toronto Metropolitan University}
  \city{Toronto}
  \country{Canada}
}

\author{Ebrahim Bagheri}
\orcid{0000-0002-5148-6237}
\affiliation{%
 \institution{University of Toronto}
 \city{Toronto}
 \country{Canada}
 }


\begin{document}

\title{\textsc{EviQE}: Evidence Selection for LLM-Based Query Expansion}

\begin{abstract}

LLM-based query expansion increasingly conditions reformulation on documents retrieved from the target corpus, yet most work focuses on how to generate expansions rather than which documents the model should read. We propose \textsc{EviQE}, which aggregates documents retrieved by multiple reformulators, selects a compact evidence set, and uses it for one grounded expansion step. This separates evidence selection from generation and treats reformulators as complementary retrieval perspectives. Across three TREC DL and five BEIR benchmarks, reformulators frequently retrieve distinct relevant documents, so pooled candidates provide higher relevant-document coverage than any individual source. The strongest gains come from relevance-based evidence selection: LLM-Score consistently outperforms direct reformulation, cold-start expansion, and single-source seeded expansion. Additional retrieval-generation rounds provide little benefit once strong conditioning evidence has been selected and can reduce effectiveness.

\end{abstract}

\maketitle
\vspace{-1em}
\section{Introduction}

% Query reformulation has been central to improving information retrieval~\cite{rocchio1971relevance,bhogal2007review,qiu1993concept} through approaches such as reducing vocabulary mismatch, exposing latent aspects of the query, or aligning queries with the target corpus. LLMs have expanded the space of reformulation strategies by generating pseudo-documents, hypothetical answers, reasoning traces, and corpus-aware rewrites to be used as retrieval surrogates or expansion text~\cite{exploringresearch,Genqr,Genqrensemble,qa-expand}. Methods such as Query2Doc~\cite{query2doc}, HyDE~\cite{gao2022precise}, CSQE~\cite{csqe}, LameR~\cite{lamer}, MuGI~\cite{mugi}, and ThinkQE~\cite{lei2025thinkqe} show LLM-generated expansions improve first-stage retrieval without modifying the retrieval method.

Query reformulation has long been central to information retrieval~\cite{rocchio1971relevance,bhogal2007review,qiu1993concept}, reducing vocabulary mismatch, exposing latent query aspects, and aligning queries with the target corpus. LLMs expand this space by generating pseudo-documents, hypothetical answers, reasoning traces, and corpus-aware rewrites as retrieval surrogates or expansion text~\cite{exploringresearch,Genqr,Genqrensemble,qa-expand}. Methods such as Query2Doc~\cite{query2doc}, HyDE~\cite{gao2022precise}, CSQE~\cite{csqe}, LameR~\cite{lamer}, MuGI~\cite{mugi}, and ThinkQE~\cite{lei2025thinkqe} show LLM expansions improve first-stage retrieval without changing the retrieval method.

% Existing expansion methods differ in prompting, outputs, and retrieval-generation rounds, but largely focus on how reformulations are generated. In corpus-aware approaches such as CSQE and LameR, generation is conditioned on documents retrieved from the target corpus~\cite{query2doc,gao2022precise,csqe,lamer}, so the reformulation depends on both the language model and the documents supplied as context. These documents constitute the \textbf{evidence} available during generation, yet comparatively little attention has been paid to how these documents are selected. Since different retrieval strategies surface different documents for the same query, expansion effectiveness may depend not only on the generated text but also on the documents used to generate it.

Existing methods differ in prompting, outputs, and retrieval-generation rounds, but largely focus on how reformulations are generated. In corpus-aware approaches like CSQE and LameR, generation is conditioned on documents retrieved from the target corpus~\cite{query2doc,gao2022precise,csqe,lamer}, so the reformulation depends on both the language model and the documents supplied as context. These documents are the evidence available during generation, yet how they are selected has received little attention. Since different retrieval strategies surface different documents for the same query, expansion effectiveness may depend not only on the generated text but on the documents used to generate it.

% Most corpus-aware expansion methods condition generation on a single set of retrieved documents obtained from the original query, a particular reformulation strategy, or a previous iteration of an expansion pipeline. Existing approaches also commonly treat reformulators as competing alternatives. AdaRewriter and ReFormeR~\cite{adarewriter,bigdeli2026reformer} explicitly select one strategy per query, even though different reformulators often retrieve different relevant documents for the same need. Iterative methods such as ThinkQE~\cite{lei2025thinkqe} repeatedly retrieve, read, and reformulate, but each round simultaneously changes the expansion and adds retrieved documents, so observed gains may stem from successive reformulations, additional documents, or both, raising whether similar gains are achievable through stronger document selection before generation. We are therefore motivated to view query expansion as a document-selection problem, rather than a pure generation one.

Most corpus-aware methods condition generation on a single retrieved set—from the original query, one reformulation strategy, or a prior pipeline iteration—and commonly treat reformulators as competing alternatives. AdaRewriter and ReFormeR~\cite{adarewriter,bigdeli2026reformer} select one strategy per query, even though different reformulators often retrieve different relevant documents for the same need. Iterative methods like ThinkQE~\cite{lei2025thinkqe} repeatedly retrieve, read, and reformulate, but each round changes both the expansion and the retrieved documents, so gains may stem from successive reformulations, added documents, or both. This motivates viewing query expansion as a document-selection problem rather than a pure generation one.

% Motivated by this perspective, we propose \textsc{EviQE}, a query expansion framework that explicitly separates document selection from expansion generation. In \textsc{EviQE}, multiple reformulators are executed independently, the documents they retrieve are aggregated into a unified candidate pool, and a compact set of informative passages is selected prior to a single grounded generation step. Unlike reformulator-selection approaches that choose a single rewriting strategy or rank-fusion approaches that combine retrieval results after ranking, \textsc{EviQE} selects the retrieved documents that will be provided to the LLM before expansion generation takes place. Rather than viewing reformulators solely as alternative query generators, \textsc{EviQE} treats them as complementary retrieval perspectives that expose different documents relevant to the same information need. The key contributions of this work are summarized as follows.

Motivated by this, we propose EviQE, which explicitly separates document selection from generation. Multiple reformulators run independently, their retrieved documents are aggregated into a unified candidate pool, and a compact set of informative passages is selected before a single grounded generation step. Unlike reformulator-selection approaches that pick one strategy or rank-fusion approaches that combine results after ranking, EviQE selects the documents provided to the LLM before generation. It treats reformulators as complementary retrieval perspectives exposing different documents relevant to the same need. Our contributions are:

\begin{itemize}[noitemsep, topsep=0pt]
\item We reformulate LLM-based query expansion as a document-selection problem and provide a framework for disentangling the contribution of conditioning documents from that of reformulation generation.

\item We propose \textsc{EviQE}, a document-centric query expansion framework that aggregates documents retrieved by multiple reformulation methods and selects a compact set of passages prior to grounded generation.

\item We provide an empirical analysis of reformulator complementarity and show that different reformulation methods frequently expose distinct relevant documents for the same information need.

\item We evaluate \textsc{EviQE} on TREC DL and BEIR benchmarks and show that aggregating and selecting retrieved documents from multiple reformulators consistently improves retrieval effectiveness while reducing the benefit of additional retrieval-generation iterations.
\end{itemize}
Our results show that different reformulation methods frequently retrieve complementary relevant documents, allowing pooled candidate sets to achieve substantially higher relevant-document coverage than any individual reformulator. Selecting passages from these pools consistently improves retrieval effectiveness relative to cold-start expansion, direct reformulation, and single-source seeded expansion.

\begin{table}[t]
\centering
\small
\setlength{\tabcolsep}{4pt}
\caption{Reformulator complementarity analysis. QCov and DocCov measure query- and document-level coverage, respectively; $\Delta_{\mathrm{pool}}$ denotes the recall gain of pooled retrieval over the best individual reformulator.}\label{tab:rq1_evidence_reservoir}
\vspace{-0.5em}
\resizebox{\columnwidth}{!}{%
{\renewcommand{\arraystretch}{0.8}
\begin{tabular}{lccc|ccccc}
\toprule
Dataset & QCov & DocCov & Jaccard
& $|D(q)|$ & BestSingle & PoolRecall & $\Delta_{\mathrm{pool}}$ & Density \\
\midrule
TREC DL 2019   & 0.964 & 0.072 & 0.342 & 36.6 & 0.172 & 0.325 & +0.153 & 0.557 \\
TREC DL 2020   & 0.951 & 0.092 & 0.406 & 32.7 & 0.223 & 0.347 & +0.124 & 0.486 \\
DL-Hard        & 0.693 & 0.088 & 0.329 & 38.7 & 0.180 & 0.285 & +0.105 & 0.229 \\
SciFact        & 0.858 & 0.826 & 0.552 & 22.6 & 0.863 & 0.921 & +0.058 & 0.050 \\
TREC-COVID     & 0.991 & 0.015 & 0.257 & 41.9 & 0.021 & 0.064 & +0.043 & 0.675 \\
FiQA           & 0.467 & 0.236 & 0.379 & 32.4 & 0.308 & 0.445 & +0.137 & 0.035 \\
DBPedia-Entity & 0.861 & 0.079 & 0.354 & 35.7 & 0.244 & 0.393 & +0.149 & 0.213 \\
TREC-News      & 0.944 & 0.101 & 0.477 & 27.1 & 0.172 & 0.281 & +0.109 & 0.436 \\
\bottomrule
\end{tabular}}}
\vspace{-1em}
\end{table}

% -----------------------------------------------------------------------------
% TABLE 2: HERO TABLE -- nDCG@10 only, all datasets x all strategies, R1.
% -----------------------------------------------------------------------------
\begin{table}[t]
\caption{Main retrieval results (\ndcg{}). Cold denotes ThinkQE seeded from the initial BM25 ranking. \best{Bold} indicates the best non-oracle result. Average rows are reported separately for TREC DL and BEIR benchmarks. Improvements of LLM-Score over Best Single and Best QR are significant on both average rows (paired per-query t-test $p<0.05$).}
\label{tab:main}
\centering
\small
\vspace{-0.5em}
\resizebox{\columnwidth}{!}{%
{\renewcommand{\arraystretch}{0.8}
\begin{tabular}{l c c c c c c c}
\toprule
 & \multicolumn{2}{c}{\textit{Direct / cold}} & \multicolumn{2}{c}{\textit{Single-source seeded}} & \multicolumn{3}{c}{\textit{Portfolio-selected documents}} \\
\cmidrule(lr){2-3} \cmidrule(lr){4-5} \cmidrule(lr){6-8}
Dataset        & Cold   & Best QR & Single mean & Best single & Freq   & RRF    & LLM-Score \\
\midrule
TREC DL 2019   & 0.6673 & 0.6798 & 0.6573 & 0.6714 & 0.6874 & 0.6895 & \best{0.6961} \\
TREC DL 2020   & 0.6321 & 0.6322 & 0.6303 & 0.6384 & 0.6313 & 0.6368 & \best{0.6603} \\
DL-Hard        & 0.3489 & 0.3470 & 0.3497 & 0.3614 & 0.3659 & 0.3597 & \best{0.3718} \\
SciFact        & 0.7376 & 0.7183 & 0.7373 & 0.7400 & 0.7352 & 0.7363 & \best{0.7435} \\
TREC-COVID     & 0.7494 & 0.7423 & 0.7515 & 0.7711 & 0.7572 & 0.7511 & \best{0.7760} \\
FiQA           & 0.2575 & 0.2460 & 0.2560 & 0.2585 & 0.2536 & 0.2568 & \best{0.2785} \\
DBPedia-Entity & 0.4264 & 0.3987 & 0.4283 & 0.4314 & 0.4287 & 0.4273 & \best{0.4410} \\
TREC-News      & 0.5004 & 0.4778 & 0.4990 & 0.5082 & 0.4901 & 0.4924 & \best{0.5174} \\
\midrule
\textbf{DL Avg.}   & 0.5494 & 0.5530 & 0.5458 & 0.5571 & 0.5615 & 0.5620 & \best{0.5760} \\
\textbf{BEIR Avg.} & 0.5343 & 0.5166 & 0.5344 & 0.5419 & 0.5330 & 0.5328 & \best{0.5513} \\
\bottomrule
\end{tabular}}}
\vspace{-1em}

\end{table}
\vspace{-1em}
%=======================================================================
\section{Methodology}
\label{sec:method}

% \textsc{EviQE} is motivated by the observation that different reformulation methods often retrieve different relevant documents for the same information need. Existing query expansion methods typically use a single set of retrieved documents to condition generation, even though alternative reformulation strategies may uncover additional relevant information. \textsc{EviQE} formulates LLM-based query expansion as an inference-time document-selection problem.

\textsc{EviQE} is motivated by the observation that different reformulation methods retrieve different relevant documents for the same need, yet existing query expansion methods condition generation on a single retrieved set. \textsc{EviQE} formulates LLM-based query expansion as an inference-time document-selection problem. Given a query $q$, corpus $C$, fixed retriever $R$, and fixed generator $G$, let $S(q)$ denote the set of retrieved documents supplied to the generator. The generator produces an expansion $e=G(q,S(q))$, which is appended to the original query to form the expanded query $\hat q=q\oplus e$. The expanded query is then submitted to the retriever, yielding the final ranking $L(\hat q)=R(\hat q;C)$. The objective is to identify the set of retrieved documents that, when provided to the generator, produces the most effective expanded query.

\begin{equation}
S^{\star}(q)
=
\arg\max_{S\in\Omega_B(q)}
\mathcal{M}
\!\left(
R(q\oplus G(q,S);C),
y^{\star}(q)
\right),
\end{equation}

where $\Omega_B(q)$ denotes the family of candidate subsets drawn from the retrieved-document pool, $\mathcal{M}$ is the retrieval metric, and $y^{\star}(q)$ denotes the unknown relevance labels associated with query $q$.

% Since $y^{\star}(q)$ is unavailable at inference time, the optimal document set cannot be computed directly. \textsc{EviQE} approximates $S^{\star}(q)$ by first constructing a candidate pool from the documents retrieved by multiple reformulation strategies and then selecting a compact subset to guide generation. This formulation separates document selection from expansion generation. Whereas most LLM-based query expansion methods implicitly determine the documents available to the generator through a particular retrieval or reformulation strategy, \textsc{EviQE} explicitly models and optimizes the selection of those documents while keeping the generation process fixed.

Since $y^{\star}(q)$ is unavailable at inference time, \textsc{EviQE} approximates $S^{\star}(q)$ by constructing a candidate pool from multiple reformulator probes and selecting a compact subset before generation. This separates two decisions that are usually coupled in LLM-based expansion, namely what the generator reads and how it rewrites. We keep the generator fixed and vary only the selected documents.

\vspace{-0.5em}

\subsection{The \textsc{EviQE} Framework}
\label{sec:pipeline}

To approximate the optimal document set, \textsc{EviQE} constructs a candidate pool of retrieved documents from multiple reformulation strategies and selects a compact subset to condition generation.

Let $\mathcal{P}=\{m_0,m_1,\ldots,m_M\}$ denote a portfolio of reformulators, where $m_0(q)=q$ corresponds to the original query. Each reformulator produces a query variant $q_i=m_i(q)$, which is independently submitted to the retriever to obtain a top-$K$ ranking,
\begin{equation}
L_i(q)=\operatorname{TopK}(R(q_i;C)).
\end{equation}
Rather than selecting a single reformulation strategy, \textsc{EviQE} retains the documents retrieved by all reformulators. The resulting candidate pool is defined as
\begin{equation}
D(q)
=
\bigcup_{i=0}^{M} L_i(q).
\end{equation}
The union operation produces a de-duplicated set of candidate documents while preserving source-specific rank information. Since different reformulators frequently retrieve different relevant documents, the resulting candidate pool often provides broader coverage of the information need than any individual reformulator.

Because the candidate pool is typically larger than the amount of information that can be supplied to the language model, \textsc{EviQE} subsequently performs document selection. Given a selector $A$ with scoring function $a_A(q,d)$, the framework selects the top-$B$ documents according to:
\begin{equation}
S_A(q)
=
\operatorname*{Top\text{-}B}_{d\in D(q)}
a_A(q,d),
\end{equation}
\noindent where $B$ denotes the document budget. All candidate documents are scored with respect to the original query rather than the reformulated query that retrieved them. The selected document set $S_A(q)$ is then supplied to a fixed generator, which produces an expansion:
\begin{equation}
e = G(q,S_A(q)).
\end{equation}
The generated expansion is appended to the original query and submitted to the retriever for final ranking. Because the retriever, generator, prompt, and document budget remain fixed across selectors, differences in retrieval effectiveness can be attributed directly to the quality of the selected documents.

\vspace{-0.5em}
\subsection{Evidence Selection Strategies}
\label{sec:strategies}

\noindent Since the effectiveness of \textsc{EviQE} depends on the quality of the selected evidence, we instantiate $a_A$ with three evidence-selection strategies that prioritize different signals of document utility: agreement, rank position, and estimated relevance. These let us examine whether useful expansion evidence is primarily characterized by repeated discovery across reformulators, strong retrieval rankings, or semantic relevance to the original query.

\noindent \textbf{Agreement-based Selection.} Documents are scored by cross-source agreement, assigning higher scores to those retrieved by more reformulators, under the assumption that documents repeatedly retrieved across reformulation perspectives are more likely broadly relevant to the information need. Ties are broken by the best rank assigned by any reformulator.

\noindent \textbf{Rank-based Selection.} This strategy uses Reciprocal Rank Fusion (RRF)~\cite{rrf} to combine cross-source agreement with rank position, so documents retrieved by multiple reformulators and ranked near the top score highest. The rationale is that rank provides a utility signal beyond agreement alone, prioritizing highly ranked evidence even when reformulator overlap is limited.

\noindent \textbf{Relevance-based Selection.} This strategy directly estimates relevance using an LLM judge that assigns a graded relevance score~\cite{faggioli2023perspectives,upadhyay2024umbrela,thomas2024gpt,macavaney2023onesizefits,sun2023chatgpt,qin2024large} to each pooled document with respect to the original query. Unlike the previous strategies, which use retrieval behavior as a proxy for relevance, it explicitly evaluates the semantic relationship between document and information need. Since the pooled set is de-duplicated, each unique document is evaluated only once.

% \begin{table}[t]
% \centering
% \small
% \setlength{\tabcolsep}{4pt}
% \caption{Candidate pool ablation. }\label{tab:candidate_pool_ablation}
% \resizebox{\columnwidth}{!}{%
% % {\renewcommand{\arraystretch}{0.8}
% \begin{tabular}{llccc|ccccc|cc}
% \toprule
% Methods & Candidate pool & DL19 & DL20 & DL-H & SciFact & COVID & FiQA & DBPed & News & DL Avg. & BEIR Avg. \\
% \midrule
% % 1 & BM25@10 & 0.667 & 0.632 & 0.349 & 0.738 & 0.749 & 0.258 & 0.426 & 0.500 & 0.549 & 0.534 \\
% 1 & BM25@100 & 0.664 & 0.618 & 0.337 & 0.747 & 0.751 & 0.266 & 0.417 & 0.495 & 0.540 & 0.535 \\
% 4 & Original, CSQE, LameR, MuGI & 0.693 & 0.658 & 0.377 & 0.749 & 0.760 & 0.271 & 0.438 & 0.509 & 0.576 & 0.545 \\
% 6 & + Query2Doc (CoT, ZS) & 0.705 & 0.665 & 0.372 & 0.741 & 0.757 & 0.273 & 0.440 & 0.511 & 0.581 & 0.545 \\
% 8 & + QA-expand, Query2Doc (FS) & 0.698 & 0.657 & 0.376 & 0.746 & 0.765 & 0.275 & 0.441 & 0.519 & 0.577 & 0.549 \\
% 11 & + GenQR, GenQR Ensemble, Query2E & 0.696 & 0.660 & 0.372 & 0.744 & 0.776 & 0.279 & 0.441 & 0.517 & 0.576 & 0.551 \\
% \bottomrule
% \end{tabular}}
% \vspace{-1em}
% \end{table}

\begin{table}[t]
\centering
\scriptsize 
\setlength{\tabcolsep}{4pt}
\caption{Candidate pool ablation.}
\vspace{-1em}

\label{tab:candidate_pool_ablation}
{\renewcommand{\arraystretch}{0.9}
% --- Part (a): TREC DL ---
\begin{tabular}{ll cccc}
\toprule
\# & Candidate pool & DL19 & DL20 & DL-H & DL Avg. \\
\midrule
1  & BM25@100 + LLM-score                        & 0.664 & 0.618 & 0.337 & 0.540 \\
4  & Original, CSQE, LameR, MuGI     & 0.693 & 0.658 & 0.377 & 0.576 \\
6  & \;+ Query2Doc (CoT, ZS)         & 0.705 & 0.665 & 0.372 & \textbf{0.581} \\
8  & \;+ QA-expand, Query2Doc (FS)   & 0.698 & 0.657 & 0.376 & 0.577 \\
11 & \;+ GenQR, GenQR Ens., Query2E  & 0.696 & 0.660 & 0.372 & 0.576 \\
\bottomrule
\end{tabular}

\vspace{0.8em}

% --- Part (b): BEIR ---
\begin{tabular}{ll ccccc c}
\toprule
\# & Candidate pool & SciFact & COVID & FiQA & DBPed & News & BEIR Avg. \\
\midrule
1  & BM25@100 + LLM-score                      & 0.747 & 0.751 & 0.266 & 0.417 & 0.495 & 0.535 \\
4  & Original, CSQE, LameR, MuGI     & 0.749 & 0.760 & 0.271 & 0.438 & 0.509 & 0.545 \\
6  & \;+ Query2Doc (CoT, ZS)         & 0.741 & 0.757 & 0.273 & 0.440 & 0.511 & 0.545 \\
8  & \;+ QA-expand, Query2Doc (FS)   & 0.746 & 0.765 & 0.275 & 0.441 & 0.519 & 0.549 \\
11 & \;+ GenQR, GenQR Ens., Query2E  & 0.744 & 0.776 & 0.279 & 0.441 & 0.517 & \textbf{0.551} \\
\bottomrule
\end{tabular}}
\vspace{-3em}

\end{table}

\vspace{-1em}

\section{Experimental Setup}

\noindent\textbf{Datasets and evaluation.} We use TREC DL 2019, DL 2020, DL-Hard~\cite{TREC2019,TREC2020,DL_HARD} and five BEIR collections~\cite{kamalloo}. Following prior reformulation work~\cite{mugi,query2doc,lamer,csqe,exploringresearch}, \ndcg{} is adopted as the primary metric; other metrics reported on our repository.

% , except where BEIR rerun files do not provide MAP annotations. Statistical significance, where applicable, is measured using paired $t$-tests on per-query \ndcg{} differences with Bonferroni correction across compared strategies.

% counter-argument retrieval (ArguAna),

\noindent \textbf{Large Language Models.}
Our framework uses LLMs in two roles, namely as reformulators/generators producing query expansions, and as judges estimating document relevance. Unless noted, all reformulation and expansion experiments use \texttt{Qwen2.5-7B-Instruct}~\cite{qwen7b}. We additionally use \texttt{DeepSeek-V3}~\cite{deepseekv3} and \texttt{Llama-3.3-70B}~\cite{llama3} as generators, with the judge held fixed as robustness check. For document scoring, we use the UMBRELA relevance-judging framework~\cite{upadhyay2024umbrela}; following recent reproducibility findings~\cite{farzi2025umbrela}, we adopt \texttt{DeepSeek-V3} as the default judge for its strong agreement with proprietary assessors on TREC DL.

\noindent \textbf{Implementation details.}
All methods use the same BM25 retriever, document index, generator, expansion prompt, and final retrieval pipeline. Each reformulated query retrieves the top $K=10$ passages, and the selector chooses the top $B=10$ passages for generation. The portfolio contains the identity query and ten reformulation sources, namely CSQE, LameR, MuGI, Query2Doc-ZS, Query2Doc-FS, Query2Doc-CoT, QA-Expand, GenQR, GenQR-Ensemble, and Query2E. Each generator call produces two expansion candidates, appended to the original query using the same formatting for all methods~\cite{lei2025thinkqe}. Compared with a $T$-round ThinkQE pipeline, \textsc{EviQE} uses one generation round, $|\mathcal{P}|$ parallel retrievals, and up to $|D(q)|$ judge calls. For the judge, we employ graded relevance assessments ($y \in \{0, 1, 2, 3\}$). All experimental results are averaged over three independent runs. Code, runs, prompts, and detailed results are released anonymously\footnote{\url{https://github.com/queryreform-judge/EviQE}}.

% \noindent \textbf{Implementation and reproducibility details.}
% All methods use the same BM25 first-stage retriever, the same document index, the same generator, the same expansion prompt, and the same final retrieval pipeline. Unless otherwise stated, each reformulated query retrieves the top $K=10$ passages. The reformulator portfolio contains the identity query and ten reformulation sources obtained from CSQE, LameR, MuGI, Query2Doc-ZS, Query2Doc-FS, Query2Doc-CoT, QA-Expand, GenQR, GenQR-Ensemble, Query2E . The selector then chooses the top $B=10$ passages as the conditioning document set. Each generator call produces two expansion candidates, which are appended to the original query using the same formatting for all methods~\cite{lei2025thinkqe}. Unlike the $T$-round ThinkQE pipeline ($T$ sequential generation rounds, each with retrieval), \textsc{EviQE} uses one generation round, $|\mathcal{P}|$ parallel retrievals, and up to $|D(q)|$ judge calls. For the judge, we employ graded relevance assessments ($y \in \{0, 1, 2, 3\}$). All experimental results reported in this paper are the average of 3 independent runs.  We release the full set of code, hyperparameters, runs, evaluation metrics, and detailed per-dataset, per-method results for all benchmarks in our anonymous Github~\footnote{\url{https://github.com/queryreform-judge/EviQE}}.

% We also report with \texttt{Llama-3.1-8B-Instruct}~\cite{llama3} for generalizability

\noindent \textbf{Baselines.}
We compare our work against three families of methods. (i) \textbf{Direct reformulation}: each of the ten reformulators is issued directly to BM25 without iterative expansion; Best QR reports the strongest reformulator per dataset. (ii) \textbf{Iterative expansion}: ThinkQE, a standard iterative baseline, initialized from the BM25 ranking of the original query without portfolio evidence. (iii) \textbf{Single-source seeded expansion}: each reformulator independently seeds the same ThinkQE-style generator; we report Single-mean (average across sources) and Best single (best source per dataset).
% (iv) \textbf{Portfolio-seeded expansion}: the three EviQE variants, i.e., Frequency, RRF~\cite{rrf}, and LLM-Score, using the same candidate pool and generator, differing only in document selection.

% Per-query paired t-tests (Table 2) show LLM-Score significantly improves over Best single on 4/8 datasets (DL19, DL20, FiQA, DBPedia-Entity) and over Best QR on 4/8 (SciFact, FiQA, DBPedia-Entity, TREC-News), with both DL and BEIR average-row gains over Best single significant and the BEIR average-row gain over Best QR significant. The asymmetry is informative: Best QR is a strong baseline on DL but weak cross-domain (Table 2), so LLM-Score's advantage is significant against Best single on DL and against Best QR on BEIR. The largest effects appear on FiQA (p < 10⁻⁶ against both baselines), the lowest-density pool, indicating the gain comes from relevance selection rather than pooling volume.

\begin{figure}[t]
\centering
\includegraphics[width=0.9\columnwidth]{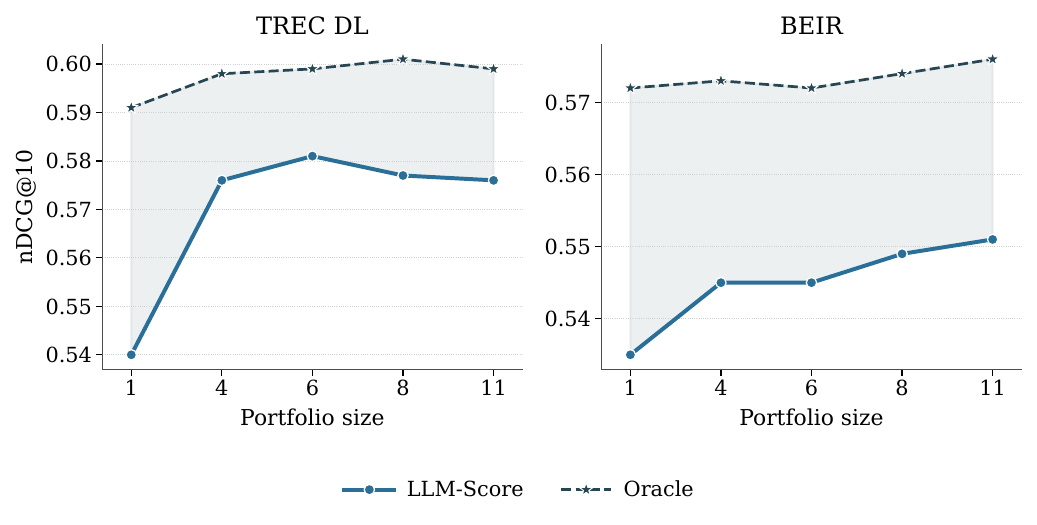}
\vspace{-1em}
\caption{Effect of portfolio size on retrieval effectiveness (\ndcg{}), averaged over TREC DL (left) and BEIR (right).}\label{fig:pool_size}
\vspace{-1em}

\end{figure}
\vspace{-1em}
%=======================================================================
\section{Results and Findings}
%=======================================================================

Our experiments are organized around 3 main research questions: \textbf{RQ1.} Do different reformulation methods retrieve the same relevant documents, or do they expose complementary documents for the same information need?
\textbf{RQ2.} Does selecting documents from a reformulator portfolio improve query expansion effectiveness, and are any gains attributable to document selection rather than a stronger reformulator or a larger candidate pool?
\textbf{RQ3.} Once high-quality conditioning documents have been identified, do additional retrieval-generation rounds provide further benefit?

\noindent \textbf{RQ1: Do reformulators retrieve complementary documents?}

RQ1 examines the degree of complementarity among the documents retrieved by different reformulation methods, asking whether reformulators agree at the query level while diverging at the document level. If they improve largely the same queries but surface different passages, selecting a single reformulator discards useful information; if they retrieve the same documents, aggregation offers little benefit. We make three observations. 

\textbf{First}, reformulators frequently converge at the query level while diverging at the document level. As shown in Table~\ref{tab:rq1_evidence_reservoir}, query hit rates are consistently high (0.964 on TREC DL 2019, 0.951 on DL 2020, 0.991 on TREC-COVID, 0.944 on TREC-News), yet per-source document coverage is far lower (0.072, 0.092, 0.015, 0.101). Their complementarity thus arises primarily from the documents they retrieve rather than the queries they address. \textbf{Second}, pooling documents across reformulators consistently exposes more relevant material than any single strategy, improving on the strongest individual source on every benchmark, with absolute recall gains from +0.043 (TREC-COVID) to +0.153 (DL 2019); BEIR shows similar trends (+0.137 on FiQA, +0.149 on DBPedia-Entity). \textbf{Third}, the pooled set is not uniformly informative, i.e., the de-duplicated pool averages 22.6–41.9 passages per query with considerable density variation (dense on TREC-COVID, sparse on FiQA and SciFact). Because the LLM can consume only a limited number of passages, aggregation alone is insufficient and expansion effectiveness also depends on identifying the most informative passages from the pool, not merely their quantity.

\begin{table}[t]
\centering
\small
\setlength{\tabcolsep}{4pt}
\caption{Model robustness on TREC DL under \texttt{DeepSeek-V3} and \texttt{Llama-3.3-70B} as generator.}
\vspace{-1em}
\resizebox{0.9\columnwidth}{!}{%
\label{tab:model_robustness}
{\renewcommand{\arraystretch}{0.8}
\begin{tabular}{llcccc}
\toprule
Model & Strategy & DL19 & DL20 & DL-H & DL Avg. \\
\midrule
\multirow{5}{*}{\texttt{DeepSeek-V3}}
& Best QR & \best{0.709} & 0.648 & 0.364 & \underline{0.574} \\
% & ThinkQE & 0.675 & \best{0.651} & 0.365 & 0.564 \\
& ThinkQE & 0.662 & 0.641 & 0.362 & 0.555 \\
& Frequency & 0.679 & 0.638 & \underline{0.368} & 0.562 \\
& RRF & 0.675 & 0.638 & 0.356 & 0.556 \\
& LLM-Score & \underline{0.700} & \underline{0.651} & \best{0.373} & \best{0.575} \\
\midrule
\multirow{5}{*}{\texttt{Llama-3.3-70B}}
& Best QR & 0.671 & \best{0.663} & 0.367 & \underline{0.567} \\
% & ThinkQE & 0.680 & 0.645 & \best{0.375} & 0.566 \\
% & ThinkQE & 0.662 & 0.641 & 0.362 & 0.555 \\
& ThinkQE & 0.674 & 0.630 & \underline{0.370} & 0.558 \\
& Frequency & 0.681 & 0.638 & 0.369 & 0.563 \\
& RRF & \underline{0.683} & 0.632 & \underline{0.372} & 0.562 \\
& LLM-Score & \best{0.688} & \underline{0.651} & 0.370 & \best{0.570} \\
\bottomrule
\end{tabular}}}
\vspace{-2em}
\end{table}

% \subsection{RQ2: Does \textsc{EviQE} evidence improve one-shot expansion?}
\noindent \textbf{RQ2: Where do document-selection gains come from?}

We compare \textsc{EviQE} against increasingly strong baselines that isolate whether gains arise from a stronger reformulated query, a stronger individual source, pooling alone, or the combination of pooling and selection in \textsc{EviQE}. Table~\ref{tab:main} reports results across all eight benchmarks, supporting the following observations.

\textbf{First}, portfolio-based selection consistently improves one-shot expansion. Relative to ThinkQE, \textsc{EviQE} with relevance-based selection raises the DL average from 0.5494 to 0.5760 (+4.8\%) and the BEIR average from 0.5343 to 0.5513 (+3.2\%), with gains on all three DL and all five BEIR benchmarks, despite a single expansion round and unchanged retriever, generator, prompt, and document budget.
\textbf{Second}, the gains are not explained by a better reformulated query alone. The Best QR baseline, controlling for the strongest reformulation issued directly to BM25, trails \textsc{EviQE} on both DL (0.5530 vs.\ 0.5760) and BEIR (0.5166 vs.\ 0.5513). The benefit comes from using reformulators as complementary retrieval perspectives whose documents guide expansion.
\textbf{Third}, the gains are not from selecting a strong individual source. The Single-mean baseline ties Cold on BEIR (0.5344 vs.\ 0.5343), and even Best Single, assuming oracle access to the strongest source, remains below \textsc{EviQE} on DL (0.5571 vs.\ 0.5760) and BEIR (0.5419 vs.\ 0.5513). 
\textbf{Fourth}, selection is critical once a pool of documents is available. On BEIR, agreement- and rank-based selection (0.5330, 0.5328) sit only marginally above Cold, whereas relevance-based selection reaches 0.5513; the largest gains come from identifying useful documents within the pool, not pooling alone.
\textbf{Fifth}, effectiveness depends on both the selector and the candidate pool. Table~\ref{tab:candidate_pool_ablation} and Figure~\ref{fig:pool_size} show the same relevance-based selector applied to the original query alone (BM25@100) reaches only 0.540/0.535 on DL/BEIR, versus 0.576/0.551 when drawing from the reformulator portfolio. Much of the gain is recovered with small portfolios, while larger ones add robustness; a gap to the oracle (using qrels as documents) remains across all sizes, suggesting future gains lie in better selection rather than larger pools.
\textbf{Finally}, the pattern is robust across generation backbones. Table~\ref{tab:model_robustness} shows relevance-based selection achieves the strongest average under both \texttt{DeepSeek-V3} and \texttt{Llama-3.3-70B}.

\begin{figure}[t]
\centering
\includegraphics[width=\columnwidth]{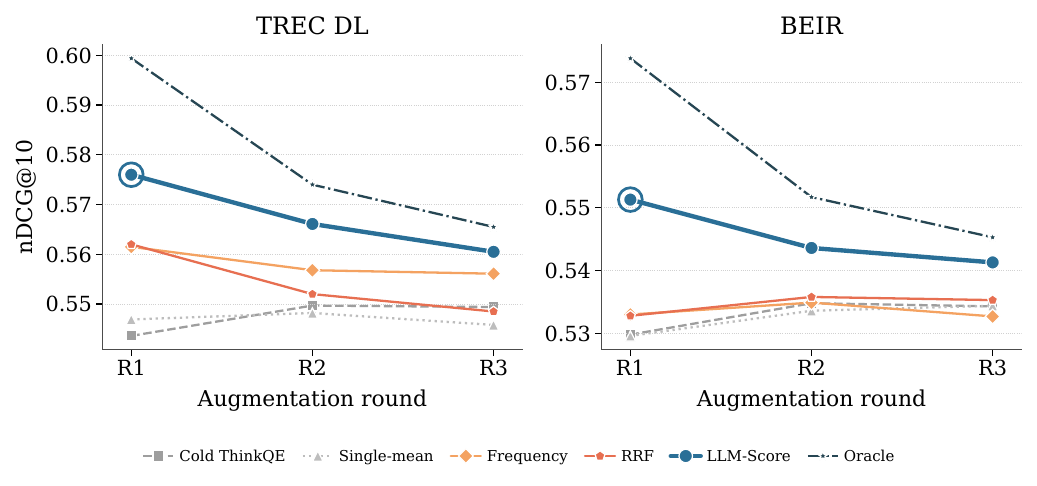}
\vspace{-2.5em}
\caption{Iteration dynamics across 3 rounds, averaged
nDCG@10 over TREC DL (left) and BEIR (right).}
\label{fig:iter_summary}
\vspace{-2em}

\end{figure}

\noindent \textbf{RQ3: Do additional retrieval-generation rounds provide further benefit?} We examine whether iterative retrieval-generation pipelines still help once high-quality conditioning documents have been identified. Many recent expansion methods derive effectiveness from multiple rounds, yet it is unclear whether these gains come from repeated reformulation or simply from improved access to useful documents.

% To answer RQ3, we examine whether iterative retrieval-generation pipelines continue to provide benefits once high-quality conditioning documents have already been identified. This question is important because many recent query expansion methods derive their effectiveness from multiple retrieval-generation rounds, yet it remains unclear whether these gains arise from repeated reformulation or simply from improved access to useful conditioning documents.
We observe that \textbf{first}, strong document selection consistently achieves its best performance in the first round. Relevance-based selection reaches its highest BEIR-averaged \ndcg{} after a single expansion round (0.5513), before declining to 0.5436 and 0.5413 in the second and third rounds, respectively. The same pattern is observed on the DL average and across individual datasets. Notably, a single expansion round conditioned on carefully selected documents outperforms multiple rounds of iterative expansion operating over weaker document sets. The oracle pool follows a similar but steeper trajectory, suggesting iterative retrieval mainly recovers documents strong selection has already identified rather than uncovering new information.
\textbf{Second}, iterative expansion helps little when the initial document set is weak. On BEIR, Cold ThinkQE improves only marginally across rounds (0.5298, 0.5348, 0.5343), and Single-mean is similarly flat, while agreement- and rank-based selection stay flat or decline. These results indicate the primary benefit of iteration is acquiring better conditioning documents; once high-quality documents are supplied, further cycles add little and may introduce noise that degrades effectiveness.

% \textbf{Second}, iterative expansion provides only limited benefits when the initial document set is weak. On BEIR, Cold ThinkQE improves only marginally from 0.5298 in the first round to 0.5348 and 0.5343 in the second and third rounds, respectively, while the Single-mean baseline exhibits a similarly flat trend. Agreement-based and rank-based selection also remain largely unchanged or decline slightly across rounds. Taken together, these results suggest that the primary benefit of iterative expansion arises from acquiring better conditioning documents. Once high-quality documents have been identified and supplied to the generator, additional retrieval-generation cycles provide little benefit and may introduce noise that degrades retrieval effectiveness.

\vspace{-1em}

%=======================================================================
\section{Concluding Remarks}
%=======================================================================

This paper examined the role of conditioning documents in LLM-based query expansion. We showed that different reformulation methods retrieve distinct relevant documents, and proposed \textsc{EviQE}, which separates document selection from generation through portfolio-based pooling and selection. Across TREC DL and BEIR, selecting high-quality conditioning documents consistently improves effectiveness over direct reformulation, cold-start expansion, and single-source seeding, while reducing the benefit of additional iterations. These findings indicate that document selection is a critical yet underexplored component of query expansion, and that much of the benefit attributed to iterative pipelines may stem from better conditioning documents rather than repeated reformulation.

\section*{GenAI Usage Disclosure}

Large language models were used for light editing of author-written text, including grammar correction, improving fluency. All text reflects the authors' own ideas; no sections were produced entirely by a generative model. GenAI tools were also used to assist with coding tasks, all code was reviewed and validated by the authors.

\balance

\bibliographystyle{ACM-Reference-Format}
\bibliography{sample-base}

@String{Computing = "Computing" }

@String{Computer = "{IEEE} Computer" }

@String{Springer = "Springer-Verlag" }

@inproceedings{csqe,
    title = "Corpus-Steered Query Expansion with Large Language Models",
    author = "Lei, Yibin  and
      Cao, Yu  and
      Zhou, Tianyi  and
      Shen, Tao  and
      Yates, Andrew",
    editor = "Graham, Yvette  and
      Purver, Matthew",
    booktitle = "Proceedings of the 18th Conference of the European Chapter of the Association for Computational Linguistics (Volume 2: Short Papers)",
    month = mar,
    year = "2024",
    address = "St. Julian{'}s, Malta",
    publisher = "Association for Computational Linguistics",
    url = "https://aclanthology.org/2024.eacl-short.34/",
    doi = "10.18653/v1/2024.eacl-short.34",
    pages = "393--401"
}

@article{query2doc,
  title={Query2doc: Query expansion with large language models},
  author={Wang, Liang and Yang, Nan and Wei, Furu},
  journal={arXiv preprint arXiv:2303.07678},
  year={2023}
}

@inproceedings{Genqrensemble,
  title={Genqrensemble: Zero-shot llm ensemble prompting for generative query reformulation},
  author={Dhole, Kaustubh D and Agichtein, Eugene},
  booktitle={European Conference on Information Retrieval},
  pages={326--335},
  year={2024},
  organization={Springer}
}

@article{Genqr,
  title={Generative query reformulation for effective adhoc search},
  author={Wang, Xiao and MacAvaney, Sean and Macdonald, Craig and Ounis, Iadh},
  journal={arXiv preprint arXiv:2308.00415},
  year={2023}
}

@article{exploringresearch,
  title={Query expansion by prompting large language models},
  author={Jagerman, Rolf and Zhuang, Honglei and Qin, Zhen and Wang, Xuanhui and Bendersky, Michael},
  journal={arXiv preprint arXiv:2305.03653},
  year={2023}
}

@inproceedings{mugi,
    title = "Exploring the Best Practices of Query Expansion with Large Language Models",
    author = "Zhang, Le  and
      Wu, Yihong  and
      Yang, Qian  and
      Nie, Jian-Yun",
    editor = "Al-Onaizan, Yaser  and
      Bansal, Mohit  and
      Chen, Yun-Nung",
    booktitle = "Findings of the Association for Computational Linguistics: EMNLP 2024",
    month = nov,
    year = "2024",
    address = "Miami, Florida, USA",
    publisher = "Association for Computational Linguistics",
    url = "https://aclanthology.org/2024.findings-emnlp.103/",
    doi = "10.18653/v1/2024.findings-emnlp.103",
    pages = "1872--1883",
}

@inproceedings{lamer,
    title = "Retrieval-Augmented Retrieval: Large Language Models are Strong Zero-Shot Retriever",
    author = "Shen, Tao  and
      Long, Guodong  and
      Geng, Xiubo  and
      Tao, Chongyang  and
      Lei, Yibin  and
      Zhou, Tianyi  and
      Blumenstein, Michael  and
      Jiang, Daxin",
    editor = "Ku, Lun-Wei  and
      Martins, Andre  and
      Srikumar, Vivek",
    booktitle = "Findings of the Association for Computational Linguistics: ACL 2024",
    month = aug,
    year = "2024",
    address = "Bangkok, Thailand",
    publisher = "Association for Computational Linguistics",
    url = "https://aclanthology.org/2024.findings-acl.943/",
    doi = "10.18653/v1/2024.findings-acl.943",
    pages = "15933--15946"
}

@article{qa-expand,
  title={QA-Expand: Multi-Question Answer Generation for Enhanced Query Expansion in Information Retrieval},
  author={Seo, Wonduk and Lee, Seunghyun},
  journal={arXiv preprint arXiv:2502.08557},
  year={2025}
}

@inproceedings{kamalloo,
    author = {Kamalloo, Ehsan and Thakur, Nandan and Lassance, Carlos and Ma, Xueguang and Yang, Jheng-Hong and Lin, Jimmy},
    title = {Resources for Brewing BEIR: Reproducible Reference Models and Statistical Analyses},
    year = {2024},
    isbn = {9798400704314},
    publisher = {Association for Computing Machinery},
    address = {New York, NY, USA},
    url = {https://doi.org/10.1145/3626772.3657862},
    doi = {10.1145/3626772.3657862},
    booktitle = {Proceedings of the 47th International ACM SIGIR Conference on Research and Development in Information Retrieval},
    pages = {1431–1440},
    numpages = {10},
    location = {Washington DC, USA},
    series = {SIGIR '24}
}

@article{TREC2019,
  title={Overview of the TREC 2019 deep learning track},
  author={Craswell, Nick and Mitra, Bhaskar and Yilmaz, Emine and Campos, Daniel and Voorhees, Ellen M},
  journal={arXiv preprint arXiv:2003.07820},
  year={2020}
}

@article{TREC2020,
  author       = {Nick Craswell and
                  Bhaskar Mitra and
                  Emine Yilmaz and
                  Daniel Campos},
  title        = {Overview of the {TREC} 2020 deep learning track},
  journal      = {CoRR},
  volume       = {abs/2102.07662},
  year         = {2021},
  url          = {https://arxiv.org/abs/2102.07662},
  eprinttype    = {arXiv},
  eprint       = {2102.07662},
  bibsource    = {dblp computer science bibliography, https://dblp.org}
}

@inproceedings{DL_HARD,
  title={How deep is your learning: The DL-HARD annotated deep learning dataset},
  author={Mackie, Iain and Dalton, Jeffrey and Yates, Andrew},
  booktitle={Proceedings of the 44th International ACM SIGIR Conference on Research and Development in Information Retrieval},
  pages={2335--2341},
  year={2021}
}

@article{rocchio1971relevance,
  title={Relevance feedback in information retrieval},
  author={Rocchio Jr, Joseph John},
  journal={The SMART retrieval system: experiments in automatic document processing},
  year={1971},
  publisher={Englewood Cliffs}
}

@article{bhogal2007review,
  title={A review of ontology based query expansion},
  author={Bhogal, Jagdev and MacFarlane, Andrew and Smith, Peter},
  journal={Information processing \& management},
  volume={43},
  number={4},
  pages={866--886},
  year={2007},
  publisher={Elsevier}
}

@inproceedings{qiu1993concept,
  title={Concept based query expansion},
  author={Qiu, Yonggang and Frei, Hans-Peter},
  booktitle={Proceedings of the 16th annual international ACM SIGIR conference on Research and development in information retrieval},
  pages={160--169},
  year={1993}
}

@inproceedings{farzi2025umbrela,
author = {Farzi, Naghmeh and Dietz, Laura},
title = {Does UMBRELA work on other LLMs?},
year = {2025},
isbn = {9798400715921},
publisher = {Association for Computing Machinery},
address = {New York, NY, USA},
url = {https://doi.org/10.1145/3726302.3730317},
doi = {10.1145/3726302.3730317},
booktitle = {Proceedings of the 48th International ACM SIGIR Conference on Research and Development in Information Retrieval},
pages = {3214–3222},
numpages = {9},
location = {Padua, Italy},
series = {SIGIR '25}
}

@misc{llama3,
      title={The Llama 3 Herd of Models}, 
      author={Aaron Grattafiori et al},
      year={2024},
      eprint={2407.21783},
      archivePrefix={arXiv},
      primaryClass={cs.AI},
      url={https://arxiv.org/abs/2407.21783}, 
}

@misc{qwen7b,
      title={Qwen Technical Report}, 
      author={Jinze Bai and Shuai Bai and Yunfei Chu and Zeyu Cui and Kai Dang and Xiaodong Deng and Yang Fan and Wenbin Ge and Yu Han and Fei Huang and Binyuan Hui and Luo Ji and Mei Li and Junyang Lin and Runji Lin and Dayiheng Liu and Gao Liu and Chengqiang Lu and Keming Lu and Jianxin Ma and Rui Men and Xingzhang Ren and Xuancheng Ren and Chuanqi Tan and Sinan Tan and Jianhong Tu and Peng Wang and Shijie Wang and Wei Wang and Shengguang Wu and Benfeng Xu and Jin Xu and An Yang and Hao Yang and Jian Yang and Shusheng Yang and Yang Yao and Bowen Yu and Hongyi Yuan and Zheng Yuan and Jianwei Zhang and Xingxuan Zhang and Yichang Zhang and Zhenru Zhang and Chang Zhou and Jingren Zhou and Xiaohuan Zhou and Tianhang Zhu},
      year={2023},
      eprint={2309.16609},
      archivePrefix={arXiv},
      primaryClass={cs.CL},
      url={https://arxiv.org/abs/2309.16609}, 
}

@misc{deepseekv3,
      title={DeepSeek-V3 Technical Report}, 
      author={DeepSeek-AI and Aixin Liu et al},
      year={2025},
      eprint={2412.19437},
      archivePrefix={arXiv},
      primaryClass={cs.CL},
      url={https://arxiv.org/abs/2412.19437}, 
}

@inproceedings{rrf,
author = {Cormack, Gordon V. and Clarke, Charles L A and Buettcher, Stefan},
title = {Reciprocal rank fusion outperforms condorcet and individual rank learning methods},
year = {2009},
isbn = {9781605584836},
publisher = {Association for Computing Machinery},
address = {New York, NY, USA},
url = {https://doi.org/10.1145/1571941.1572114},
doi = {10.1145/1571941.1572114},
booktitle = {Proceedings of the 32nd International ACM SIGIR Conference on Research and Development in Information Retrieval},
pages = {758–759},
numpages = {2},
location = {Boston, MA, USA},
series = {SIGIR '09}
}

@ARTICLE{upadhyay2024umbrela,
  title   = {UMBRELA: UMbrela is the (Open-Source Reproduction of the) Bing RELevance Assessor},
  author  = {Shivani Upadhyay and Ronak Pradeep and Nandan Thakur and Nick Craswell and Jimmy Lin},
  year    = {2024},
  journal = {arXiv:2406.06519}
}

@misc{gao2022precise,
      title={Precise Zero-Shot Dense Retrieval without Relevance Labels}, 
      author={Luyu Gao and Xueguang Ma and Jimmy Lin and Jamie Callan},
      year={2022},
      eprint={2212.10496},
      archivePrefix={arXiv},
      primaryClass={cs.IR},
      url={https://arxiv.org/abs/2212.10496}, 
}

@inproceedings{faggioli2023perspectives,
  author    = {Faggioli, Guglielmo and others},
  title     = {Perspectives on Large Language Models for Relevance Judgment},
  booktitle = {Proceedings of the ACM SIGIR International Conference on Theory
               of Information Retrieval (ICTIR)},
  year      = {2023},
  publisher = {ACM},
}

@inproceedings{thomas2024gpt,
  author    = {Thomas, Paul and Spielman, Seth and Craswell, Nick and Mitra, Bhaskar},
  title     = {{LLM} Judges for Relevance Grading in Information Retrieval},
  booktitle = {Proceedings of the 47th International ACM SIGIR Conference on
               Research and Development in Information Retrieval (SIGIR)},
  year      = {2024},
  publisher = {ACM},
}

@inproceedings{macavaney2023onesizefits,
  author    = {MacAvaney, Sean and Cohan, Arman},
  title     = {One-Shot Labeling for Automatic Relevance Estimation},
  booktitle = {Proceedings of the 46th International ACM SIGIR Conference on
               Research and Development in Information Retrieval (SIGIR)},
  year      = {2023},
  publisher = {ACM},
}

@inproceedings{sun2023chatgpt,
  author    = {Sun, Weiwei and others},
  title     = {Is {ChatGPT} Good at Search? Investigating Large Language Models
               as Re-Ranking Agents},
  booktitle = {Proceedings of the 2023 Conference on Empirical Methods in
               Natural Language Processing (EMNLP)},
  year      = {2023},
  publisher = {Association for Computational Linguistics},
}

@inproceedings{qin2024large,
  author    = {Qin, Zhen and others},
  title     = {Large Language Models are Effective Text Rankers with Pairwise Ranking Prompting},
  booktitle = {Findings of the Association for Computational Linguistics: {NAACL} 2024},
  year      = {2024},
  publisher = {Association for Computational Linguistics},
}

@misc{adarewriter,
      title={AdaRewriter: Unleashing the Power of Prompting-based Conversational Query Reformulation via Test-Time Adaptation}, 
      author={Yilong Lai and Jialong Wu and Zhenglin Wang and Deyu Zhou},
      year={2025},
      eprint={2506.01381},
      archivePrefix={arXiv},
      primaryClass={cs.CL},
      url={https://arxiv.org/abs/2506.01381}, 
}

@inbook{bigdeli2026reformer,
   title={ReFormeR: Learning and Applying Explicit Query Reformulation Patterns},
   ISBN={9783032213006},
   ISSN={1611-3349},
   url={http://dx.doi.org/10.1007/978-3-032-21300-6_30},
   DOI={10.1007/978-3-032-21300-6_30},
   booktitle={Advances in Information Retrieval},
   publisher={Springer Nature Switzerland},
   author={Bigdeli, Amin and Incesu, Mert and Arabzadeh, Negar and Clarke, Charles L. A. and Bagheri, Ebrahim},
   year={2026},
   pages={400–408} }

@inproceedings{lei2025thinkqe,
    title = "{T}hink{QE}: Query Expansion via an Evolving Thinking Process",
    author = "Lei, Yibin  and
      Shen, Tao  and
      Yates, Andrew",
    editor = "Christodoulopoulos, Christos  and
      Chakraborty, Tanmoy  and
      Rose, Carolyn  and
      Peng, Violet",
    booktitle = "Findings of the Association for Computational Linguistics: EMNLP 2025",
    month = nov,
    year = "2025",
    address = "Suzhou, China",
    publisher = "Association for Computational Linguistics",
    url = "https://aclanthology.org/2025.findings-emnlp.965/",
    doi = "10.18653/v1/2025.findings-emnlp.965",
    pages = "17772--17781",
    ISBN = "979-8-89176-335-7"
}


\end{document}